\documentclass[aps,prb,twocolumn,superscriptaddress,showpacs]{revtex4}
\usepackage{graphicx}
\usepackage{mathrsfs}
\usepackage{bm}
\usepackage{amsmath}
\usepackage{dcolumn}
\usepackage{epstopdf}
\usepackage{dsfont}
\usepackage{amssymb}
\usepackage{tabularx}
\usepackage{array}
\usepackage{colordvi}

\begin{document}
\title{Quantum Anomalous Hall Effect in Atomic Crystal Layers from In-Plane Magnetization}
\author{Yafei Ren}
\affiliation{ICQD, Hefei National Laboratory for Physical Sciences at Microscale, and Synergetic Innovation Center of Quantum Information and Quantum Physics, University of Science and Technology of China, Hefei, Anhui 230026, China}
\affiliation{CAS Key Laboratory of Strongly-Coupled Quantum Matter Physics and Department of Physics, University of Science and Technology of China, Hefei, Anhui 230026, China}
\author{Junjie Zeng}
\affiliation{ICQD, Hefei National Laboratory for Physical Sciences at Microscale, and Synergetic Innovation Center of Quantum Information and Quantum Physics, University of Science and Technology of China, Hefei, Anhui 230026, China}
\affiliation{CAS Key Laboratory of Strongly-Coupled Quantum Matter Physics and Department of Physics, University of Science and Technology of China, Hefei, Anhui 230026, China}
\author{Xinzhou Deng}
\affiliation{ICQD, Hefei National Laboratory for Physical Sciences at Microscale, and Synergetic Innovation Center of Quantum Information and Quantum Physics, University of Science and Technology of China, Hefei, Anhui 230026, China}
\affiliation{CAS Key Laboratory of Strongly-Coupled Quantum Matter Physics and Department of Physics, University of Science and Technology of China, Hefei, Anhui 230026, China}
\author{Fei Yang}
\affiliation{Department of Physics, Beihang University, Beijing 100191, China}
\author{Hui Pan}
\email[Correspondence author:~]{hpan@buaa.edu.cn}
\affiliation{Department of Physics, Beihang University, Beijing 100191, China}
\author{Zhenhua Qiao}
\email[Correspondence author:~]{qiao@ustc.edu.cn}
\affiliation{ICQD, Hefei National Laboratory for Physical Sciences at Microscale, and Synergetic Innovation Center of Quantum Information and Quantum Physics, University of Science and Technology of China, Hefei, Anhui 230026, China}
\affiliation{CAS Key Laboratory of Strongly-Coupled Quantum Matter Physics and Department of Physics, University of Science and Technology of China, Hefei, Anhui 230026, China}
\date{\today}

\begin{abstract}
  We theoretically report that, with \textit{in-plane} magnetization, the quantum anomalous Hall effect (QAHE) can be realized in two-dimensional atomic crystal layers with preserved inversion symmetry but broken out-of-plane mirror reflection symmetry. We take the honeycomb lattice as an example, where we find that the low-buckled structure, which makes the system satisfy the symmetric criteria, is crucial to induce QAHE. The topologically nontrivial bulk gap carrying a Chern number of $\mathcal{C}=\pm1$ opens in the vicinity of the saddle points $M$, where the band dispersion exhibits strong anisotropy. We further show that the QAHE with electrically tunable Chern number can be achieved in Bernal-stacked multilayer systems, and the applied interlayer potential differences can dramatically decrease the critical magnetization to make the QAHE experimentally feasible.
\end{abstract}

\maketitle
\section{Introduction}
Quantum anomalous Hall effect (QAHE), manifesting itself as quantized Hall conductance and vanishing longitudinal conductance, has attracted broad interests recently.\cite{rev_2D_TopoPhase} In analogy to the quantum Hall effect from strong out-of-plane magnetic field, the QAHE has been intensively studied by introducing out-of-plane ferromagnetism in various systems, such as three-dimensional (3D) topological insulator thin films,\cite{rev_2D_TopoPhase, QAHE_MagTI_FangZh_10,QAHE_MagTI_exp_XueQK_13, QAHE_MagTIexp_WangKL_14, QAHE_MagTI_exp_Checkelsky_14, QAHE_MagTI_exp_Moodera_15} quantum-well based structures\cite{QAHE_QW_InPlane_LiuCX_13, QAHE_QW_ZhangCW_11} and atomic crystal layers, e.g. honeycomb-lattice systems.\cite{rev_2D_TopoPhase, QAHE_G_Qiao_10, QAHE_G_Qiao_11, QAHE_G_Qiao_12, QAHE_Si_Ezawa_12, QAHE_QSHE_QVHE_Si_Ezawa_13, QAHE_QVHE_Si_Yao_14}
Experimentally, by employing ferromagnetic insulating substrates, AHE has been reported in graphene though much efforts are still required to realize the quantized version.\cite{QAHE_G_AFM_Qiao_14, QAHE_G_ZhangJ_15, LMO_ZengCG_14, QAHE_G_AFM_exp_Shi_15} For such systems, a perpendicular magnetic field is usually required to align the magnetization of the system that prefers the in-plane orientation. This inspired us to think whether it is possible to realize QAHE by using in-plane magnetization. To the best of our knowledge, except limited studies in quantum well based structures,\cite{QAHE_QW_InPlane_LiuCX_13, QAHE_QW_ZhangCW_11} the QAHE from in-plane magnetization has not been reported in the 2D atomic crystal layers.

By using of symmetry analysis, we investigate the possibility of realizing QAHE from in-plane magnetization in atomic crystal layers and show that the QAHE can occur in systems with preserved inversion symmetry but breaking out-of-plane mirror reflection symmetry. We numerically verified that the QAHE cannot appear in a planar honeycomb lattice, e.g. graphene, but can be realized in \textit{low-buckled} honeycomb-lattice systems like silicene. With an in-plane magnetization of proper orientation, the topologically nontrivial bulk gap hosting the QAHE with a Chern number of $\mathcal{C}=\pm1$ opens up around the $M$ points, the saddle points with strong anisotropy, which is completely different from other models of QAHE from out-of-plane magnetization with band gap opening around the isotropic Dirac points.\cite{rev_2D_TopoPhase, Haldane_88, QAHE_G_Qiao_10, QAHE_Si_Ezawa_12, QAHE_QSHE_QVHE_Si_Ezawa_13, QAHE_QVHE_Si_Yao_14, QAHE_QW_ZhangCW_11} We further show that, for QAHE in Bernal-stacked multilayer systems, the Chern number of higher values can be achieved and can be tuned via electrical means. The applying of interlayer potential differences dramatically decrease the lowest critical magnetization strength to make the QAHE experimentally feasible.

\begin{figure}
  \includegraphics[width=9 cm]{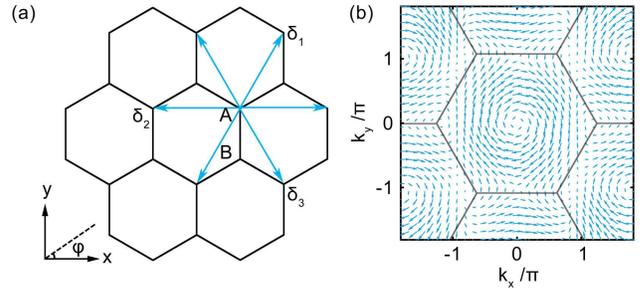}
  \caption{ (color online) (a) Schematic of the top view of low-buckled honeycomb lattice. A and B denote the A/B sublattices while $\pm \delta_i$ ($i=1$-$3$) are the next-nearest-neighbor sites of the site A. $\phi$ indicates the orientation of the in-plane magnetization. (b) Local distribution of the intrinsic-Rashba SOC $(\lambda_{\rm{Rx}}, \lambda_{\rm{Ry}})$ in momentum space. }
  \label{Contour1}
\end{figure}

This paper is organized as follows. In Sec.\,II, we display the model of our calculation. Section\,III shows the possibility of realizing QAHE by in-plane magnetization by symmetric analysis. The numerical results of monolayer and multilayer systems are presented in Sec.\,IV and Sec.\,V, respectively. We summarize in Sec.\,VI.

\section{Model}
For the monolayer low-buckled honeycomb lattice with in-plane magnetization, the general $\pi$-band tight-binding Hamiltonian can be written as following,\cite{QAHE_QVHE_Si_Yao_14,QAHE_Si_Ezawa_12}
\begin{equation}\label{EQSingleH}
H=H_{0}+H_{\rm{I}}+H_{\rm{IR}}+H_{\rm{M}}+H_{AB}+H_{\rm{ER}},
\end{equation}
where
\begin{eqnarray*}
&& H_{0}       =  -t\sum_{\langle ij \rangle}c^\dagger_{i}c_{j}, \\
&& H_{\rm{I}}    =  {\mathrm{i}} t_{\rm{I}}\sum_{\langle\langle ij \rangle\rangle}\nu_{ij}c^\dagger_{i}{s}_{z}c_{j}, \\
&& H_{\rm{IR}} = -{\mathrm{i}} t_{\rm{IR}}\sum_{\langle\langle ij \rangle\rangle}\mu_{ij} c^\dagger_{i}(\bm{s} \times \hat{\bm{d}}_{ij})_{z} c_{j}, \\
&& H_{\rm{M}} = \lambda \sum_{i}c^\dagger_{i}\hat{\bm{m}}\cdot \bm{s} \, c_{i}, \\
&& H_{AB}=\Delta \sum_{i\in A}c^\dagger_{i}c_{i}- \Delta \sum_{i\in B}c^\dagger_{i}c_{i}, \\
&& H_{\rm{ER}}= {\mathrm{i}} t_{\mathrm{ER}} \sum_{\langle{ij}\rangle} c^\dagger_{i} (\bm{s}{\times}\hat{\bm{d}}_{ij}) {\cdot}
\hat{\bm{z}} \,  c_{j}.
\end{eqnarray*}
Here, $c^\dagger_{i}=(c^\dagger_{i\uparrow},c^\dagger_{i\downarrow})^{\rm{T}}$ is the creation operator for an electron at the $i$-th site with $\uparrow$ and $\downarrow$ representing the spin up and down states. The first term $H_0$ stands for the nearest-neighbor hopping with an amplitude of $t$ while the second term $H_{\rm{I}}$ is the intrinsic spin-orbit coupling (SOC) of honeycomb lattice, where $\nu_{ij}={\bm{d}_i \times \bm{d}_j}/{|\bm{d}_i \times \bm{d}_j|}$ with $\bm{d}_{i,j}$ being two nearest bonds connecting the next-nearest neighbor sites. These two terms correspond to the Hamiltonian of a planar honeycomb lattice like graphene with out-of-plane mirror reflection symmetry (i.e. $z \rightarrow -z$).\cite{GrapheneTB} This symmetry can be broken by the low-buckled structure, which is reflected by the intrinsic-Rashba SOC $H_{\rm{IR}}$ displayed as the third term where $\mu_{ij}=\pm 1$ for A/B sublattices, $\bm{s}$ are spin-Pauli matrices, and $\hat{\bm{d}}_{ij}$ is a unit vector pointing from site $j$ to $i$. The orientation and strength of the effective spin-orbit field of intrinsic-Rashba SOC in the reciprocal space have been displayed in Fig.~\ref{Contour1}b where one can find that the SOC strength vanishes at $M$, $K/K'$, and $\Gamma$ points [See more details in Appendix A].

The first three terms correspond to the pristine low-buckled honeycomb lattice while the last three ones can be applied externally.
$H_{\rm{M}}$ represents the in-plane magnetization, with the strength and orientation being separately $\lambda$ and $\hat{\bm{m}}=(\cos{\varphi},\sin{\varphi},0)$ as displayed in Fig.~\ref{Contour1}a. $H_{AB}$ stands for the staggered sublattice potential while the last term $H_{\rm{ER}}$ is the extrinsic-Rashba SOC comes from the structural inversion asymmetry, which, different from the intrinsic one, breaks not only the out-of-plane mirror reflection symmetry but also the inversion symmetry.

For the Bernal-stacked bilayer system, the tight-binding Hamiltonian can be expressed as
\begin{align}
\label{EQ:BilayerH}
H_{\rm{BL}} &= H^{\rm{T}}+H^{\rm{B}}+t_{\perp}\sum_{i\in {\rm{T}},j\in {\rm{B}}}(c^\dagger_{i}c_{j} +H.c.) \\ \nonumber
& + U/2 \sum_{i\in \rm{T}}c^\dagger_{i}c_{i}-U/2 \sum_{i\in \rm{B}}c^\dagger_{i}c_{i},
\end{align}
where $H^{\rm{T(B)}}$ denotes the Hamiltonian for the top (bottom) monolayer low-buckled honeycomb lattice. $t_{\perp}$ is the interlayer hopping energy between `dimer' sites, i.e. the two atomic sites with the atom at lower layer being directly below that at the upper layer.\cite{rev_BLG_McCann_13} And $U$ corresponds to the interlayer potential difference from out-of-plane electric field. For Bernal-stacked multilayer systems, the Hamiltonian can be obtained similarly with only the interlayer hopping between `dimer' sites being included.

\begin{table}
  \caption{Parity of in-plane (out-of-plane) magnetization $H_{\parallel}$ ($H_{\perp}$), intrinsic (extrinsic) Rashba SOC $H_{\rm{IR}}$ ($H_{\rm{ER}}$), staggered AB sublattice potentials $H_{AB}$, velocity $\bm{v}$, momentum $\bm{k}$, and electric field $\bm{E}$ under the symmetric operations of time reversal $\mathcal{T}$, out-of-plane mirror reflection $\mathcal{M}_z$, and inversion $\mathcal{I}$. $+/-$ indicates even/odd parity. }
  \label{Tab1}
    \begin{tabular}{c|cc|cc|c|ccc}
  \hline
                                        & $~H_{\parallel}~$ & $~H_{\perp}~$ & $~H_{\rm{IR}}~$  & $~H_{\rm{ER}}~$ & $~H_{AB}~$ & $~~\bm{v}~~$  & $~~\bm{k}~~$ & $~~\bm{E}~~$ \\ \hline
  $\mathcal{T}$             &   $-$                    &   $-$               & +                           & +                       & +     & $-$                           & $-$                       & +    \\ \hline
  $~~\mathcal{M}_z~~$        &   $-$                    &   +                  & $-$                        & $-$                     & +     & +                           & +                       & +    \\ \hline
  $\mathcal{I}$              &   +                       &   +                  & +                           & $-$                        & $-$   & $-$                           & $-$                       & $-$     \\
  \hline
    \end{tabular}
\end{table}

\section{Symmetry Analysis}
We begin from symmetry analysis of the Hamiltonian and the corresponding Berry curvature $\bm{\Omega}_n(\bm{k}) = \Omega^z_n(\bm{k}) \hat{z}$ based on the anomalous velocity in the presence of in-plane electric field $\bm{E}$,\cite{rev_BerryCuv}
\begin{align}
\label{BerryCuv}
\bm{v}_n(\bm{k})=\frac{\partial \varepsilon_n(\bm{k})}{\hbar\partial \bm{k}}-\frac{e}{\hbar}\bm{E}\times \bm{\Omega}_n(\bm{k}).
\end{align}
The integration of $\Omega^z_n(\bm{k})$ over the first Brillouin zone is Chern number that characterizes the topological property of the $n$-th band.\cite{Thouless,NiuQ} We focus on the operations of inversion $\mathcal{I}$, time reversal $\mathcal{T}$, and out-of-plane mirror reflection $\mathcal{M}_z$, i.e. $z \rightarrow -z$. Under these operations, the parities of velocity $\bm{v}$, momentum $\bm{k}$, and electric field $\bm{E}$ are listed in Tab.~\ref{Tab1}. We first consider a planar honeycomb lattice with vanishing intrinsic-Rashba SOC, e.g. graphene, which is invariant under these three operations. In the presence of in-plane magnetization, the tight-binding Hamiltonian reads
\begin{equation}\label{H1G}
H_1=H_{0}+H_{\rm{I}}+H_{\parallel}
\end{equation}
where $H_{\parallel}$ corresponds to the magnetization term $H_{\rm{M}}$ with $\parallel$ being employed to emphasize its in-plane orientation.
\begin{figure*}
  \includegraphics[width=16 cm]{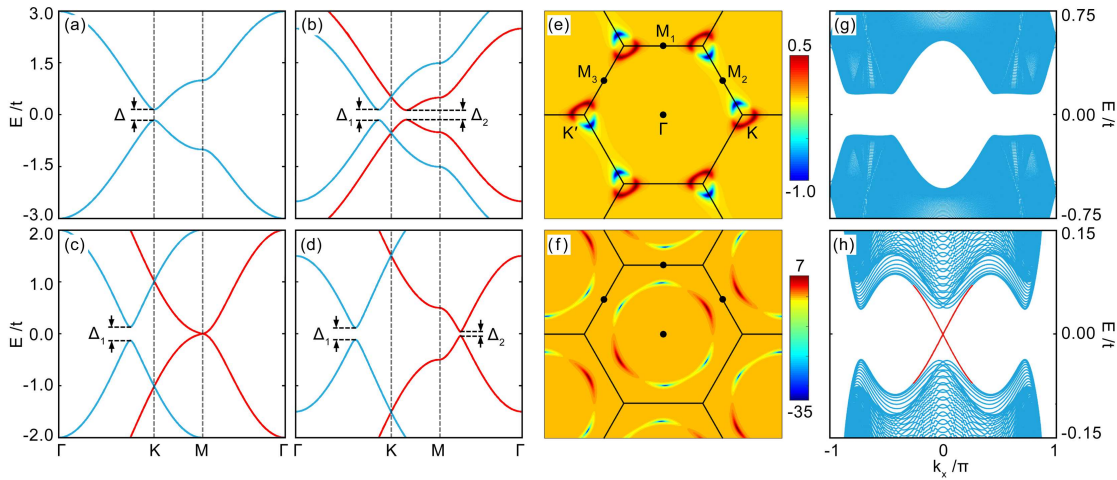}
  \caption{ (color online) (a-d): Band structures of low-buckled honeycomb lattice in the presence of different in-plane magnetization strengths of $\lambda/t=0.0$ (a), 0.5 (b), 1.0 (c) and 1.5 (d) at the orientation of $\phi=\pi/6$. With the increase of $\lambda$, a topological phase transition occurs accompanying with a bulk band gap closing (c) and reopening (d). (e) and (f): Berry curvature distribution in the Brillouin zone for the insulating states shown in (b) and (d). (g) and (h): The corresponding zigzag-ribbon band structures for the systems shown in (b) and (d). Red lines in (f) highlight the chiral gapless edge modes of the QAHE. In our calculations, the SOCs are chosen to be $t_{\rm{I}}=t_{\rm{IR}}=0.03t$.}
  \label{figure_bandStr}
\end{figure*}
Since the in-plane magnetization has odd parities under both $\mathcal{T}$ and $\mathcal{M}_z$ operations while has even parity under inversion $\mathcal{I}$ as shown in Tab.~\ref{Tab1}, $H_{\parallel}$ cannot break the invariance of the system under the joint operation of $\mathcal{T}\otimes\mathcal{M}_z\otimes\mathcal{I}$, which makes $\Omega^z_n(\bm{k})$ vanishes over the first Brilliouin zone as shown below.

Under the operation of $\mathcal{T}\otimes\mathcal{M}_z\otimes\mathcal{I}$, one can find that the left-handed side of Eq.~\eqref{BerryCuv} is invariant while the other side changes to $\frac{\partial \varepsilon_n(\bm{k})}{\hbar\partial \bm{k}}-\frac{e}{\hbar}(-\bm{E})\times \bm{\Omega}_n(\bm{k})$ since only the in-plane electric field has odd parity under the joint operation of $\mathcal{T}\otimes\mathcal{M}_z\otimes\mathcal{I}$. The invariance of the system under this symmetry requires that the
\begin{align}
\frac{\partial \varepsilon_n(\bm{k})}{\hbar\partial \bm{k}}-\frac{e}{\hbar}\bm{E}\times \bm{\Omega}_n(\bm{k})= \frac{\partial \varepsilon_n(\bm{k})}{\hbar\partial (\bm{k})}-\frac{e}{\hbar}(-\bm{E})\times \bm{\Omega}_n(\bm{k}).
\end{align}
Therefore, the Berry curvature mush vanish over the whole first Brillouin zone and thus the Chern number is zero. This conclusion is consistent with the analysis on the in-plane anti-ferromagnetism induced Berry curvature.\cite{ChenHua}

In order to induce nonzero Berry curvature, the joint symmetry of $\mathcal{T}\otimes\mathcal{M}_z\otimes\mathcal{I}$ in graphene must be broken. One possible method is to break the inversion symmetry $\mathcal{I}$ by introducing staggered sublattice potential term $H_{AB}$. With this term, however, the system is still invariant under the operation of $\mathcal{T}\otimes\mathcal{M}_z$. Through a similar analysis, one can find that this symmetry guarantees the oddness of the Berry curvature $\Omega^z_n(\bm{k})$ as function of momentum $\bm{k}$. Therefore, the Chern number obtained by integrating $\Omega^z_n(\bm{k})$ over the first Brillouin zone vanishes. Up to now, one can conclude that, in order to induce nonzero Chern number, the symmetries of the system under both $\mathcal{T}\otimes\mathcal{M}_z\otimes\mathcal{I}$ and $\mathcal{T}\otimes\mathcal{M}_z$ must be broken simultaneously.

Fortunately, this symmetric criteria can be met by introducing, instead of $H_{AB}$, intrinsic-Rashba SOC from the low-buckled structure, which is odd under $\mathcal{M}_z$ while is even under inversion $\mathcal{I}$. Thus, its combination with in-plane magnetization as shown in Eq.~\eqref{EQSingleH} breaks both symmetries of $\mathcal{T}\otimes\mathcal{M}_z$ and $\mathcal{T}\otimes\mathcal{M}_z\otimes\mathcal{I}$ leading to nonzero Berry curvature that is even function of momentum guaranteed by the inversion symmetry. Therefore, nonzero Chern number is possible, corresponding to QAHE in insulator and AHE in metal.

In addition to the intrinsic-Rashba SOC, the extrinsic-Rashba SOC $H_{\rm{ER}}$ from structural inversion asymmetry, e.g. from substrate, also has odd parity under out-of-plane mirror reflection $\mathcal{M}_z$. However, different from the intrinsic one, it is also odd under the inversion $\mathcal{I}$ and therefore, its combination with in-plane magnetization preserves the the joint symmetry of $\mathcal{T}\otimes\mathcal{M}_z\otimes\mathcal{I}$ leading to zero Berry curvature.

In contrast to the in-plane one, the out-of-plane magnetization itself breaks both the joint symmetry of $\mathcal{T} \otimes \mathcal{M}_z \otimes \mathcal{I}$ and $\mathcal{T} \otimes \mathcal{M}_z$ simultaneously. Thus, together with either intrinsic- or extrinsic-Rashba SOC, it can lead to nonzero Berry curvature, which is also an even function of $\bm{k}$ guaranteed by the invariance of the Hamiltonian under either $\mathcal{I}$ or $\mathcal{M}_z\otimes \mathcal{I}$.\cite{QAHE_G_Qiao_10, QAHE_QVHE_Si_Yao_14} Such differences between in-plane and out-of-plane magnetizations in symmetry completely distinguish our model from the others in the literature. In the following, we numerically verify the QAHE from in-plane magnetization in low-buckled honeycomb lattice without staggered sublattice potential and extrinsic-Rashba SOC based on Eq.~\eqref{EQSingleH}.

\section{Monolayer Case}
In the following calculations, we do not include the staggered sublattice potential as well as the extrinsic-Rashba SOC unless otherwise noted. We set the nearest-neighbor hopping energy $t$ as the energy unit. Without loss of generality, the strengths of intrinsic and intrinsic-Rashba SOCs are set to be $t_{\rm{I}}=t_{\rm{IR}}=0.03t$ in all our calculations.

\subsection{Numerical results}
We first calculate the band structure via exact diagonalization of the above Hamiltonian with the magnetization orientation being $\varphi=\pi/6$. At $\lambda=0$, a band gap opens up at valleys $K/K'$ (see Fig.~\ref{figure_bandStr}a), harbouring a 2D $\mathbb{Z}_2$ topological insulator.\cite{QSHE_Si.Ge_Yao_11} When a nonzero magnetization $\lambda<t$ is applied, the doubly-degenerate bands are split into two species as highlighted in blue and red, which are characterized by different band gaps $\Delta_1$ and $\Delta_2$ as displayed in Fig.~\ref{figure_bandStr}b. However, this insulating phase with broken time-reversal symmetry is topologically trivial characterized by vanishing Chern number of $\mathcal{C}=0$ that can be obtained by integrating the Berry curvature shown in Fig.~\ref{figure_bandStr}e.\cite{Thouless,NiuQ} This can be further verified by the absence of gapless edge modes of the ribbon band structure (Fig.~\ref{figure_bandStr}g). When the magnetization reaches a critical value of $\lambda=t$, the band gap $\Delta_1$ is nearly unchanged while $\Delta_2$ is completely closed at $M$ points as displayed in Fig.~\ref{figure_bandStr}c. For even larger magnetization strength, i.e. $\lambda>t$, the degeneracies at the saddle points $M$ are lifted and the band gap $\Delta_2$ reopens, which changes the Berry-curvature distribution as shown in Fig.~\ref{figure_bandStr}f. One can find that the Berry curvature is negative for $M_1$ and $M_3$ but positive for $M_2$ giving rise to a negative Chern number of $\mathcal{C}=-1$. Moreover, in the corresponding ribbon band structure, chiral gapless edge modes emerges as displayed in Fig.~\ref{figure_bandStr}h in red. These characters indicates the formation of a QAHE.

We further consider the dependence of the topological phase on the magnetization orientation. Figure~\ref{Mono_layer_phase_diagram1}a shows the band gap as well as the topological phase of the system in $m_x$-$m_y$ plane with ($m_x$, $m_y$)=$\lambda$($\cos \varphi$, $\sin \varphi$). In this figure, the central white dot denotes the $\mathbb{Z}_2$ topological insulator phase at $\lambda=0$. For $\lambda<t$, the insulating phase of vanishing Chern number $\mathcal{C}=0$ occurs independent of magnetization orientation. The increase of magnetization induces band gap closing at $\lambda=t$ but reopening for $\lambda>t$ that hosts QAHE characterized by Chern numbers of $\mathcal{C}=\pm1$ as labelled in Fig.~\ref{Mono_layer_phase_diagram1}a. Different from the case of $\lambda<t$, the reopened band gap is strongly dependent on the magnetization orientation $\varphi$ and vanishes at $\varphi=n\pi/3$ ($n=0$-$5$) as shown by the white dashed lines separating QAHEs with opposite Chern numbers of $\mathcal{C}=\pm1$. This dependence of Chern number on $\varphi$ is consistent with the symmetry analysis in Refs.~[\onlinecite{QAHE_QW_InPlane_LiuCX_13}] and [\onlinecite{QAHE_Symm_Fang_12}].

\subsection{Low energy effective Hamiltonian}
\begin{figure}
  \includegraphics[width=8 cm]{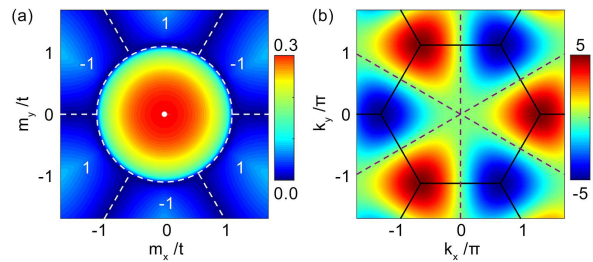}
  \caption{(a) Phase diagram of the low-buckled honeycomb lattice with $(m_x,m_y)$=$\lambda(\cos \varphi,\sin \varphi)$. The white dot at the center indicates the 2D $\mathbb{Z}_2$ topological insulator at $\lambda=0$. When $0<\lambda<t$, the system is a trivial insulator. When $\lambda>t$, the system is a QAHE with alternating Chern numbers $\mathcal{C}$=$\pm1$. Dashed lines indicate the phase boundaries. (b) Contour-plot of the amplitude of intrinsic SOC in momentum space. The first Brillouin zone is denoted by solid lines, and the intrinsic SOC vanishes along dashed lines. In our calculation, the SOCs are chosen to be $t_{\rm{I}}=t_{\rm{IR}}=0.03t$. }
  \label{Mono_layer_phase_diagram1}
\end{figure}

Here, we stress that the intrinsic-Rashba SOC from the \textit{low-buckled} structure plays an important role in the reopening of band gap $\Delta_2$. As shown in Fig.~\ref{Mono_layer_phase_diagram1}b, the amplitude of intrinsic SOC is momentum-dependent and vanishes at high-symmetric lines connecting $\Gamma$ and $M$ points as denoted by purple dashed lines. In the absence of intrinsic-Rashba SOC, this feature leads to the formation of Dirac points at $\Gamma$-$M$ lines whenever the magnetization $\lambda>t$  [See more details in Appendix B]. Thus, the QAHE cannot form without intrinsic-Rashba SOC. To better demonstrate this point, the low-energy effective Hamiltonians around $M$ points are provided as below
\begin{align}
\label{Heffs}
h_{M_1}(\bm{q},\varphi) &= +[m\sigma_z + a \sigma_x + b(\varphi) \sigma_y], \nonumber \\
h_{M_2}(\bm{q},\varphi) &= - [m\sigma_z + a \sigma_x + b(\varphi+\pi/3) \sigma_y], \\
h_{M_3}(\bm{q},\varphi) &= - [m\sigma_z + a \sigma_x + b(\varphi-\pi/3) \sigma_y], \nonumber
\end{align}
where the unit of momentum is set to be $1/\sqrt{3}a_0$ with $a_0$ being the nearest-neighbor distance. $\sigma_{x,y,z}$ are Pauli matrices and the mass term $m=-\delta\lambda+t(3q_y^2-q_x^2)/4$ shows strongly anisotropic momentum dependence with $\delta\lambda=\lambda -t$. $a=4q_xt_{\rm{I}}$ and $b=t_{\rm{IR}}(3q_y\sin \varphi-q_x\cos \varphi)$ are separately contributed from intrinsic and intrinsic-Rashba SOCs. The detailed method to obtain the effective Hamiltonian is attached in Appendix D.

After a direct diagonalization, the eigenenergies for each $M$ point are obtained to be $\varepsilon_{\pm}=\pm\sqrt{m^2+a^2+b^2}$, which is consistent with the particle-hole symmetry that assures the symmetry of band structure about the Fermi-level $\varepsilon_{\rm{F}}=0$. Thus, the band gap closing conditions can be obtained by solving the equation of $\varepsilon_{\pm}=0$. Since the three effective Hamiltonians share similar form, we take $h_{M_1}$ as an example and find that the band gap closing occurs when (i) $\delta\lambda=0$ at $(q_x,q_y)=(0,0)$ for any $\varphi$, corresponding to the critical points of topological phase transition from a trivial insulator ($\delta\lambda<0$) to a QAHE ($\delta\lambda>0$); (ii) $\delta\lambda > 0$ along the direction of $\varphi=0, \pi$ at $(q_x,q_y)=(0,\pm\sqrt{4\delta\lambda/3})$, indicating the band gaps close at $\varphi=0, \pi$ that correspond to two phase boundaries separating QAHEs with $\mathcal{C}=\pm1$. By further including the phase boundaries obtained through solving $\varepsilon_{\pm}=0$ around $M_2$ and $M_3$, one can find that there are totally six linear phase boundaries corresponding to the six dashed white lines in Fig.~\ref{Mono_layer_phase_diagram1}b. It is noteworthy that the $\varepsilon_{\pm}=0$ is always solvable for $\delta\lambda>0$ whenever $a=0$ or $b=0$. This indicates that a gapless metallic phase occurs for $\lambda>t$ in the absence of either intrinsic or intrinsic-Rashba SOC, reflecting the significance of the coexistence of these two kinds of SOCs in the formation of QAHE.

\begin{figure}
  \includegraphics[width=8 cm]{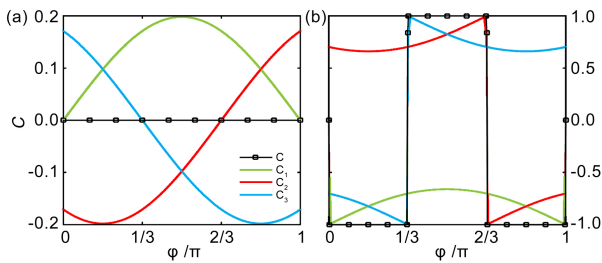}%
  \caption{Chern number $\mathcal{C}_{1,2,3}$ contributed from Berry curvature in the vicinity of $M_{1,2,3}$ points by using the low-energy effective Hamiltonian as a function of the direction of the magnetization for $\delta \lambda=-0.05t$ (a) and $0.05t$ (b). Only the Chern numbers for $0\leq \varphi \leq \pi$ are plotted for simplicity since $\mathcal{C}_i(\varphi)=-\mathcal{C}_i(\varphi+\pi)$ with $i=1$-$3$. In our calculation, the SOCs are chosen to be $t_{\rm{I}}=t_{\rm{IR}}=0.03t$. }
  \label{reduced-simplified2x2-1}
\end{figure}

To further clarify the contributions of the three $M$ points to the resulting QAHE, we illustrate the Chern number dependence on the magnetization orientation $\varphi$ for $\delta\lambda<0$ and $\delta\lambda>0$ as shown in Fig.~\ref{reduced-simplified2x2-1} by integrating the Berry curvatures contributed from $M_{1,2,3}$.\cite{Thouless,NiuQ} For $\delta\lambda<0$, the Chern numbers contributed from $M_{1,2,3}$ can be described by a sinusoidal function of $\varphi$, i.e. $\mathcal{C}_{1}=\mathcal{C}_0\sin\varphi$ and $\mathcal{C}_{{2,3}}=\mathcal{C}_0\sin(\varphi\mp 2\pi/3)$ with $\mathcal{C}_0 \simeq 0.2$ (see Fig.~\ref{reduced-simplified2x2-1}a). By summing up these analytical expressions, the total Chern number is shown to be $\mathcal{C}=\sum_i\mathcal{C}_{i}=0$, indicating a topologically trivial insulator. For $\delta\lambda > 0$, as shown in Fig.~\ref{reduced-simplified2x2-1}(b), the Chern number contributed from each $M$ point is changed approximately by $1$ or $-1$ depending on $\varphi$, leading to a nonzero total Chern number of $\mathcal{C}=\pm1$ and thus a QAHE.

\begin{figure}
  \includegraphics[width=8cm]{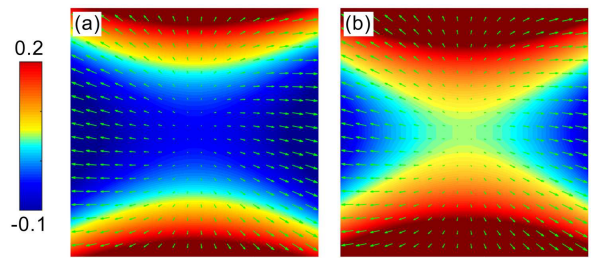}
  \caption{(color online) Pseudospin texture around $M_1$ based on effective Hamiltonian $H_{{M_1}}$ with $\varphi=\pi/6$ and magnetization strength of (a) $\lambda=0.95t$, and (b) $\lambda=1.05t$. In our calculations, the SOCs are chosen to be $t_{\rm{I}} = t_{\rm{IR}}=0.03t$.}
  \label{SpinTexture}
\end{figure}

These $2\times2$ effective Hamiltonians can be alternatively expressed as $\bm{d}\cdot \bm{\sigma}$ where $\bm{\sigma} = (\sigma_x, \sigma_y, \sigma_z)$ are the Pauli matrices and $\bm{d}=(a,b,m)$ represents the pseudospin texture~\cite{rev_2D_TopoPhase} that is intimately related to Chern number by the formula
\begin{equation}
\mathcal{C}=-\frac{1}{8\pi^2} \int d^2 k  \hat{\bm{d}} \cdot \partial_{q_x} \hat{\bm{d}} \times \partial_{q_y} \hat{\bm{d}}.
\label{CMeron}
\end{equation}
with $\hat{\bm{d}}$ being the unit vector of $\bm{d}$. In previous literatures on the quantum anomalous Hall effect
(QAHE),\cite{rev_2D_TopoPhase}
the topologically nontrivial low-energy Hamiltonian carries a Chern number of either an integer or half-integer with the corresponding pseudospin texture in momentum space being either a Skyrmion or a Meron,\cite{rev_TI_Bernevig} e.g. Haldane's model,\cite{Haldane_88} BHZ model,\cite{QAHE_MagTI_FangZh_10} and Rashba SOC based model in honeycomb lattice.\cite{QAHE_G_Qiao_10, QAHE_G_Qiao_12} However, the pseudospin texture $(a,b,m)$ in our proposed model is completely different from that of a Skyrmion or Meron, and shows strong anisotropy as plotted in Figs.~\ref{SpinTexture}a and \ref{SpinTexture}b for $\lambda<t$ and $\lambda>t$ respectively based on $h_{M_1}$.
In these figures, the arrow represents the in-plane components $(a,b)$ while the color stands for the out-of-plane component $m$. Such pseudospin texture with strong anisotropy corresponds to non-vanishing Chern number but \textit{neither an integer nor half-integer} for both $\lambda<t$ and $\lambda>t$. This character, which is intimately related to the strong anisotropic momentum dependence of $h_{M_i}$ ($i=1$-$3$), makes our low-energy model strongly distinct from previous models of realizing QAHE in the literature.\cite{rev_2D_TopoPhase, Haldane_88, QAHE_MagTI_FangZh_10, QAHE_G_Qiao_10}

\begin{figure*}
  \includegraphics[width=16 cm]{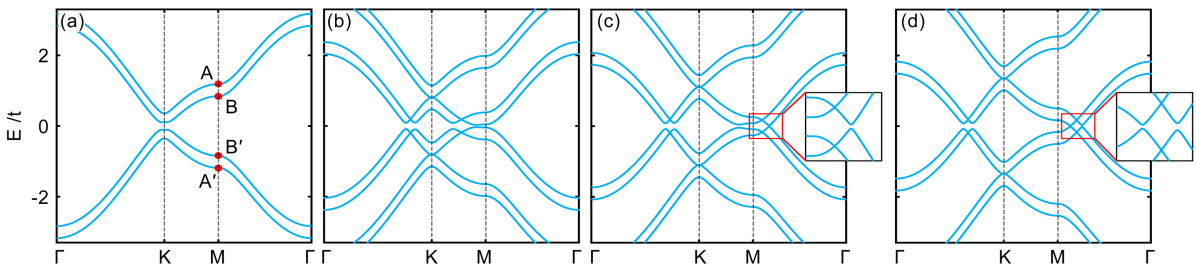}
  \caption{ (color online) Band structure evolution of chiral-stacked bilayer structure of low-buckled honeycomb lattice under an in-plane magnetization with $\varphi=\pi/6$ for different magnetization amplitudes, i.e. (a) $\lambda=0$, (b) $\lambda = 0.5t$, (c) $\lambda=1.1t$, and (d) $\lambda=1.3t$. (a) Solid red circles show the energy levels at $M$ point labelled by A, B, A$^{\prime}$, and B$^{\prime}$. Insets of (c) and (d) zoom out the band structure around $M$. In our calculation, we set intrinsic and intrinsic-Rashba SOCs as $t_{\rm{I}}= t_{\rm{IR}} = 0.03t$.}
  \label{Bi_layer_bandStr}
\end{figure*}

\subsection{Extrinsic-Rashba SOC}
In this section, we study the effect of extrinsic-Rashba SOC, which is inevitably introduced in the 2D system with a substrate that breaks the mirror-reflection symmetry. In contrast to the QAHE induced by out-of-plane magnetization where extrinsic-Rashba SOC is essential,\cite{QAHE_G_Qiao_10} our results show that the extrinsic-Rashba is detrimental to the in-plane magnetization induced QAHE.

In the presence of extrinsic-Rashba SOC, we can obtain the modified low-energy effective Hamiltonian around $M$ points by employing similar procedures mentioned in Appendix D. We take $M_1$ as an example, for which the effective Hamiltonian displayed in Eq.~\eqref{Heffs} is modified to be
\begin{eqnarray}
h^{\rm{R}}_{M_1}(\bm{q}) = u {\bm{1}}_{\sigma} + m\sigma_z + a_{\rm{R}} \sigma_x + b \sigma_y,
\end{eqnarray}
where an additional term appears with ${\bm{1}}_{\sigma}$ being the unit matrix and $u=-t_{\rm{ER}}[ \sqrt{3} q_x
\sin\varphi/2 + (1+4/\sqrt{3})q_y\cos\varphi]$. This term breaks the particle-hole symmetry of the Hamiltonian and thus the band gap is no longer symmetric about the Fermi level. Moreover, the extrinsic-Rashba SOC also changes $a$ to be $a_{\rm{R}}=a + 2t_{\rm{ER}} \sin\varphi$ while leaves $m$ and $b$ being unaffected. The modification of $a$ can shrink the bulk band gap if $a$ and $\sin \varphi$ have opposite signs. For the case of $\delta\lambda>0$ and $\varphi\neq n\pi/3$ ($n=0$-$5$), the direct band gap closes when $t_{\rm{ER}}$ satisfies the equation of $\delta\lambda=[t_{\rm{ER}} / (4t_{\rm{I}})]^2 \cos 2 \varphi$. This equation is solvable when $\cos2\varphi>0$ with the solution corresponding to the critical extrinsic-Rashba SOC strength of the phase transition from QAHE to topologically trivial insulator. If $\cos2\varphi<0$, the solution will appear around $M_2$ or $M_3$, which indicates that a topological phase transition can always be found as $t_{\rm{ER}}$ increases and thus the extrinsic-Rashba SOC is detrimental to the QAHE.

Nevertheless, we can eliminate the effect of extrinsic-Rashba SOC by sandwiching the low-buckled honeycomb lattice symmetrically between two ferromagnetic insulating substrates. The symmetric structure can not only strongly suppress the extrinsic-Rashba SOC but also enhance the proximately induced magnetization in the low-buckled honeycomb lattice.

\section{Multilayer Cases}
So far, we have verified that the in-plane magnetization induced QAHE can be formed in low-buckled honeycomb-lattice systems. However, a daunting challenge for realizing this QAHE is the extremely large magnetization that is comparable to the hopping energy $t$. Hereinbelow, we show that the lowest critical magnetization for realizing QAHE can be effectively reduced in Bernal-stacked multilayer systems. Let us first take the bilayer system as an example and adopt the same SOC parameters as those in monolayer case.

We first study the band structure evolution for different magnetization strengths $\lambda$ with $\varphi=\pi/6$. When $\lambda=0$, a topologically trivial band gap is opened at $K$ and $K^{\prime}$ points by SOCs as shown in Fig.~\ref{Bi_layer_bandStr}a. The bilayer system gives rise to two sets of conductance and valence bands with a small energy splitting. The eigenenergies at $M$ point for one set are labelled as A/A$'$ while that for the other set are labelled as B/B$'$.  Due to the inversion symmetry and Kramers degeneracy, in this figure each band is two-fold degenerate, which can be lifted in the presence of magnetization and a topologically trivial band gap appears in the presence of both SOCs as shown in Fig.~\ref{Bi_layer_bandStr}b.
When the magnetization strength increases to the regime that leads to the closing of band gap between B and B$^{\prime}$ points but leaves A and A$^{\prime}$ points untouched as shown in Fig.~\ref{Bi_layer_bandStr}c, the topological phase transition to QAHE with $\mathcal{C}=-1$ occurs. When the magnetization strength further increases to close both band gaps between B/B$^{\prime}$ points and A/A$^{\prime}$ points, the QAHE with $\mathcal{C}=-2$ appears (see Fig.~\ref{Bi_layer_bandStr}d).

\begin{figure}
  \includegraphics[width=8 cm]{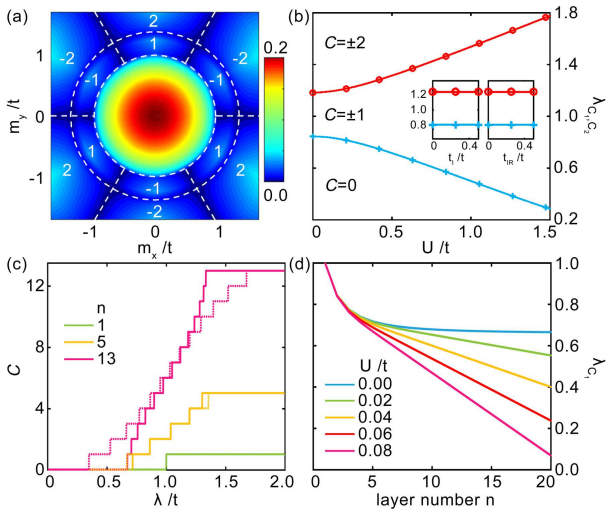}
  \caption{(a) Phase diagram of the bilayer low-buckled honeycomb-lattice in the $m_x$-$m_y$ plane. Dashed lines indicate the phase boundaries. Chern numbers are labelled in the QAHE regions accordingly.  (b) Evolution of critical phase boundaries $\lambda_{C_{1},C_{2}}$ as a function of the inter-layer potential difference.
  (c) Chern number $\mathcal{C}$ as a function of the amplitude of magnetization $\lambda$ at fixed $\varphi=-\pi/6$ for different layers. Solid and dashed curves correspond to $U=0$ and $U=0.05t$, respectively. (d) The lowest critical magnetization amplitude to induce the QAHE of $\mathcal{C}=\pm1$ as a function of the layer number $n$ at fixed $\varphi=-\pi/6$ for different potential differences. In our calculation, the SOCs are chosen to be $t_{\rm{I}}=t_{\rm{IR}}=0.03t$.}
  \label{Bi_layer_phase_diagram1}
\end{figure}

By varying $\lambda$ and $\varphi$, we calculate the bulk band gaps and the Chern numbers as shown in Fig.~\ref{Bi_layer_phase_diagram1}a, which shows the same angular dependence as that of the monolayer system. However, different from monolayer system, additional topological phases arise with Chern numbers of $\mathcal{C}=\pm2$. Moreover, topological phase transition from $\mathcal{C}=0$ to $\mathcal{C}=\pm 1$ appears at the critical magnetization $\lambda_{C_1} \simeq 0.8 t$, which is smaller than that in the monolayer case. It is noteworthy that the critical magnetization strengths to induce QAHEs with different Chern numbers are determined by the energy differences between A/A$^\prime$ or B/B$^\prime$ in the absence of magnetization, which is tunable via interlayer potential differences from electrical gating in Bernal stacked multilayer systems. This provides a possible way to further reduce the critical magnetization.

By applying an interlayer potential differences $U$, we calculated the critical magnetizations for bilayer systems as displayed in Fig.~\ref{Bi_layer_phase_diagram1}b, where one can find that the increase of $U$ can extremely decrease the critical magnetization $\lambda_{C_1}$, while correspondingly enlarges $\lambda_{C_2}$ that separates the topological phases of $\mathcal{C}=\pm1$ and $\pm2$. Therefore, the QAHE with $\mathcal{C}=\pm1$ can be achieved at a rather smaller magnetization in the presence of a considerable electric field. In addition, the dependence of $\lambda_{C_{1,2}}$ on interlayer potential difference also makes it possible to realize the QAHE with electrically tunable Chern number. Inset of Fig.~\ref{Bi_layer_phase_diagram1}b shows that the topological phase transitions are independent of the amplitudes of SOCs.

The cases for Bernal-stacked multilayer systems are similar to bilayer one. As highlighted by the solid lines in Fig.~\ref{Bi_layer_phase_diagram1}c, large-Chern-number QAHEs appear for $n$-layer systems as the magnetization $\lambda$ increases with the upper limit of $\mathcal{C}=n$. When an interlayer potential difference $U$ is applied, the critical magnetization is decreased (increased) for smaller- (larger-) Chern-number QAHEs (see the dashed lines in Fig.~\ref{Bi_layer_phase_diagram1}c). For $\lambda_{C_1}$, the lowest magnetization required to induce the QAHE with $\mathcal{C}=\pm1$, we find that it decreases slowly as the layer thickness increases for $U=0$ as shown by the light blue line of Fig.~\ref{Bi_layer_phase_diagram1}d. When the applied potential difference $U$ gradually increases, $\lambda_{C_1}$ can be dramatically decreased for multilayer systems as illustrated in Fig.~\ref{Bi_layer_phase_diagram1}d. These findings strongly suggest that a multilayer low-buckled honeycomb-lattice system under proper perpendicular electric field is a more ideal and more experimentally feasible platform to realize the QAHE from in-plane magnetization.

\section{Summary and Discussions}
In this paper, by using symmetry analysis, we have theoretically revealed the fact that, in 2D case, nonzero Chern number can only occur in systems breaking both symmetries of $\mathcal{T}\otimes\mathcal{M}_z\otimes\mathcal{I}$ and $\mathcal{T}\otimes\mathcal{M}_z$. These two symmetries are simultaneously broken by out-of-plane magnetization while are preserved for in-plane one. This makes the QAHE from in-plane magnetization can only be realized in systems with certain constrains, like in the atomic crystal layers with preserved inversion symmetry but broken out-of-plane mirror reflection symmetry. Such differences between in-plane and out-of-plane magnetizations in symmetry distinguish our model from the previous ones to realize QAHE in the literature. We numerically verified the realization of QAHE in honeycomb lattice of low-buckled structure, in which the band gap hosting QAHE occurs in the vicinity of time-reversal symmetric $M$ points where the electronic structure shows strong anisotropy and exhibits magnetization-orientation dependent non-integer Chern number.

Experimentally, the in-plane magnetization could be introduced by applying an in-plane magnetic field that cannot form Landau levels in the ultrathin films or proximately coupling with ferromagnetic insulating substrates,\cite{QAHE_G_AFM_Qiao_14, QAHE_G_ZhangJ_15} where a symmetric setup with the low-buckled honeycomb-lattice system sandwiched by two identical ferromagnetic insulating layers is required to eliminate the influence of extrinsic-Rashba SOC that is detrimental to the QAHE from in-plane magnetization.
However, the extremely large magnetization strength required, i.e. $\lambda_{C1}=t$, makes the experimental realization of the QAHE in monolayer system difficult.

There are two possible ways to overcome this difficulty. One is to decrease the nearest-neighbor hopping energy $t$ by, e.g. constructing artificial organometallic material with low-buckled honeycomb structure, where the effective nearest-neighbor hopping energy $t$ is relatively weak and exchange field is rather strong and can even be much larger than $t$.\cite{QAHE_organic_LiuF_13} The other one is to consider Bernal-stacked multilayer systems, where the critical magnetization $\lambda_{C_1}$ gradually decreases along with the increase of system thickness and could be further dramatically reduced by applying interlayer potential differences via vertical electric field. Our studies together with the recent experimental realization of Bernal-stacked multilayer silicene\cite{KHWu1, KHWu2, KHWu3, MLSi} strongly suggest the QAHE from in-plane magnetization could be experimentally achievable in such system with a vertical electric field.

Apart from the half-filled low-buckled honeycomb lattice (e.g. silicene, gemanene, and stanene), there are plenty of atomic crystal layers satisfy the symmetrical criteria discussed above, such as, organometallic materials, bismuth bilayer, black and blue phersphorene.\cite{rev_2D_TopoPhase} The QAHE from in-plane magnetization may also be realized in these systems and their hybridized structures, where the magnetization required may be small enough to be experimentally feasible. Furthermore, such symmetry analysis on Berry curvature is not limited to QAHE. With nonzero Berry curvature that is even function of momentum, the AHE is also expected in metallic system with in-plane magnetization.

\begin{acknowledgments}
We are grateful to Prof. Q.-F. Sun for valuable discussions. This work is financially supported by China Government Youth 1000-Plan Talent Program, Anhui Provincial Natural Science Foundation, National Natural Science Foundation of China (Grant No.: 11574019 and 11474265), Fundamental Research Funds for the Central Universities (Grant Nos.: WK3510000001, WK2030020027), and the National Key R \& D Program (Grant No.2016YFA0301700). The supercomputing center of USTC is gratefully acknowledged for the high-performance computing assistance.
\end{acknowledgments}

\begin{appendix}

\section{Tight-binding Hamiltonian in momentum space }
We first consider the low-buckled honeycomb lattice with in-plane magnetization and neglecting the staggered sublattice potential as well as the extrinsic-Rashba SOC. In the absence of extrinsic-Rashba SOC and staggered sublattice potential, the real-space tight-binding Hamiltonian of Eq.~\eqref{EQSingleH} can be expressed in the momentum space by doing a Fourier transform:
\begin{eqnarray}
H(\bm{k})=\left(
  \begin{array}{cccc}
    \lambda_{\rm{I}} & -\eta & \lambda_{\rm{RA}} & 0 \\
    -\eta^* & -\lambda_{\rm{I}} & 0 & \lambda_{\rm{RB}} \\
    \lambda_{\rm{RA}}^* & 0 & -\lambda_{\rm{I}} & -\eta \\
    0 & \lambda_{\rm{RB}}^* & -\eta^* & \lambda_{\rm{I}} \\
  \end{array}
\right).
\label{MLTBH}
\end{eqnarray}
Here, $\eta=t\sum_{i}\exp({\rm{i}}\bm{k}\cdot\bm{\delta^{\prime}}_i)$, $\lambda_{\rm{I}}=-2t_{\rm{I}} \sum_{i} \sin(\bm{k} \cdot \bm{\delta}_i)$ are respectively the kinetic energy and intrinsic SOC terms, where $\bm{\delta}_i$ ($i=1$-$3$) are the vectors pointing from A site to three nearest A sites as displayed in Fig.~\ref{Contour1}a while $\bm{\delta^{\prime}}_i$ ($i=1$-$3$) are the vectors pointing from the A site to its three nearest B sites.
$\lambda_{\rm{RA,RB}} = m_x-{\rm{i}}m_y \pm (\lambda_{\rm{Rx}} - {\rm{i}}\lambda_{\rm{Ry}})$ couples the spin-up and -down states at A (B) sublattice where $(m_x,m_y)=\lambda(\cos \varphi, \sin \varphi)$ and $\lambda_{\rm{Rx,Ry}} = \pm 2 t_{\rm{IR}} \sum_{i} \sin(\bm{k} \cdot \bm{\delta}_i) \hat{\delta}_{ix,iy}$ with $\hat{\delta}_{ix,iy}$ being the $x$ and $y$ components of unit vector $\bm{\delta}_i/|\bm{\delta}_i|$. Therefore, each term in $H(\bm{k})$ is momentum-dependent. For example, we have plotted the distribution of intrinsic SOC $\lambda_{\rm{I}}$ and intrinsic-Rashba SOC in the momentum space as respectively shown in Figs.~\ref{Mono_layer_phase_diagram1}b and \ref{Contour1}b in the main text. We find that the strength of intrinsic SOC vanishes along the high-symmetric lines connecting $\Gamma$ and $M$ points while that of intrinsic-Rashba SOC vanishes at $M$, $K/K'$, and $\Gamma$ points.

\begin{figure*}
  \includegraphics[width=16cm]{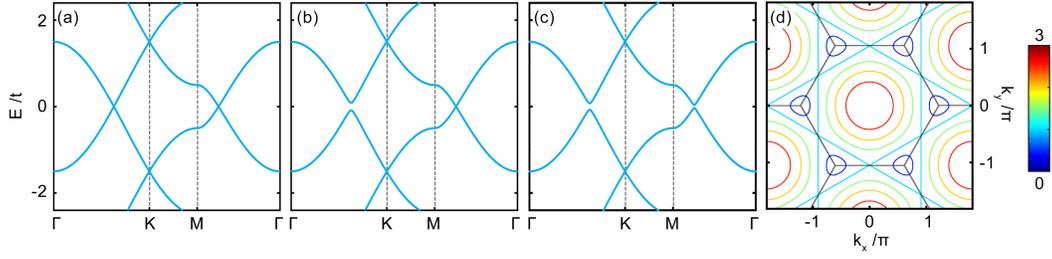}
  \caption{ (color online) (a)-(c) Band structure evolution of monolayer low-buckled system under an in-plane magnetization ($\varphi=\pi/6$ and $\lambda=1.5 t$) for different intrinsic and intrinsic-Rashba SOCs, i.e. (a) $t_{\rm{I}} = t_{\rm{IR}} = 0$, (b) $t_{\rm{I}} = 0.03t$ and $t_{\rm{IR}}=0$, and (c) $t_{\rm{I}} = t_{\rm{IR}} = 0.03t$. (d) Band crossing points in the absence of both intrinsic and intrinsic-Rashba SOCs for different magnetization strengths denoted by different colors. The straight lines linking nearest $M$ points in light blue correspond to the case with $\lambda=t$.}
  \label{BandStr}
\end{figure*}

\section{Band structure}
With the Hamiltonian in momentum space, we can study the influence of each term on the electronic structure.
In Fig.~S\ref{BandStr}d, we first display the band crossing points in the absence SOCs for different magnetization strength $\lambda$. We find that, when the magnetization strength $\lambda < t$, the band crossing points circle around $K$ and $K'$ points. When $\lambda = t$, the crossing points lie at the straight lines linking nearest $M$ points as shown by light blue lines. When $\lambda > t$, however, the crossing points circle around $\Gamma$ point rather than $K/K'$ points. Then we take $\lambda=1.5t$ as an example with the magnetization direction being $\varphi=\pi/6$ and show the band structure evolution. In the absence of both types of SOC, the band structure is displayed in Fig.~S\ref{BandStr}a, where band crossings appear. The presence of intrinsic SOC can only lift the degeneracies of crossing points away from $\Gamma$-$M$ lines while the band crossings at these lines still preserve giving rise to Dirac points (see Fig.~S\ref{BandStr}b). These Dirac points originate from the vanishing strength of intrinsic SOC along $\Gamma$-$M$ lines as displayed in Fig.~\ref{Mono_layer_phase_diagram1}b of main text. However, the presence of intrinsic-Rashba SOC can move the degeneracy points away from $\Gamma$-$M$ lines and hence a bulk band gap can be opened in the presence of both SOCs as displayed in Fig.~S\ref{BandStr}c. This bulk band gap is topologically nontrivial to host a Chern number of $\mathcal{C}=-1$ that can be calculated by using the method shown in the following.

For an insulator with breaking time-reversal symmetry, the topological property is usually characterized by Chern number that can be calculated by
\begin{equation}\label{EQ:CN}
\mathcal{C}=\frac{1}{2\pi}\sum_n\int_{\rm{BZ}}d^2\bm{k}\Omega_n,
\end{equation}
where the summation is over all occupied valence bands in the first Brillouin zone, and $\Omega_n$ is the Berry curvature for the $n$-th band in momentum space~\cite{Thouless,NiuQ} that can be expressed as:
\begin{equation}
\Omega_n(\bm{k})=-\sum_{n'\neq n}
\frac{ 2\mathrm{Im}\langle\psi_{nk}|v_x|\psi_{n'k}\rangle
\langle\psi_{n'k}|v_y|\psi_{nk}\rangle }{ (\varepsilon_{n'}-\varepsilon_{n})^{2} },
\end{equation}
where $v_{x(y)}$ is the velocity operator along $x(y)$ direction. The absolute value of $\mathcal{C}$ corresponds to the number of gapless chiral edge states along any sample boundary of the corresponding 2D system.

\section{Low-energy effective Hamiltonian}
To further analyze the influence of various SOCs, magnetization strength and orientation, we present the low-energy effective Hamiltonians around the three inequivalent $M$ points, i.e. $M_{1,2,3}$ points displayed in Fig.~\ref{figure_bandStr}e of the main text. We take $M_1$ as an example whose momentum is denoted by $\bm{k}_{M_1}$. Based on the Bloch states $\{|M_1, \bm{q}, A,\uparrow\rangle, |M_1, \bm{q}, B,\uparrow\rangle, |M_1, \bm{q}, A,\downarrow\rangle, |M_1, \bm{q}, B, \downarrow\rangle \}$ at momentum $\bm{k}= \bm{k}_{M_1}+\bm{q}$, the low-energy $4\times 4$ Hamiltonian around $M_1$ point with $\bm{k}_{M_1}\gg \bm{q}$ can be approximately written as:
\begin{eqnarray}
H_{{M_1}}(\bm{q})=\left(
  \begin{array}{cccc}
    4t_{\rm{I}}q_x & \eta_{\rm{r}} & \lambda_{\rm{RAr}} & 0 \\
    \eta_{\rm{r}}^* & -4t_{\rm{I}}q_x & 0 & \lambda_{\rm{RBr}} \\
    \lambda_{\rm{RAr}}^* & 0 & -4t_{\rm{I}}q_x & \eta_{\rm{r}} \\
    0 & \lambda_{\rm{RBr}}^* & \eta_{\rm{r}}^* & 4t_{\rm{I}}q_x \\
  \end{array}
\right)
\end{eqnarray}
where $\eta_{\rm{r}}=t(1+\frac{q_y^2-3q_x^2}{12}+\frac{2{\rm{i}}q_y}{\sqrt{3}})e^{\frac{{\rm{i}}4\pi}{3}}$, $\lambda_{\rm{RAr}}= \lambda e^{-{\rm{i}}\varphi}-t_{\rm{IR}}(3q_y-{\rm{i}}q_x)$, and $\lambda_{\rm{RBr}}= \lambda e^{-{\rm{i}}\varphi}+t_{\rm{IR}}(3q_y-{\rm{i}}q_x)$.
The unit of momentum is set to be $1/\sqrt{3}a_0$ with $a_0$ being the nearest-neighbor distance in the low-buckled honeycomb lattice. Through analyzing the basis functions under a six-folder rotation operation, we find that the basis functions in the vicinity of $M_2$ and $M_3$ can be related to that of $M_1$ via the following transformation:
\begin{widetext}
\begin{eqnarray*}
\left(
  \begin{array}{c}
    | M_2, \bm{q}, A, \uparrow \rangle \\
    | M_2, \bm{q}, B, \uparrow \rangle \\
    | M_2, \bm{q}, A, \downarrow \rangle \\
    | M_2, \bm{q}, B, \downarrow \rangle \\
  \end{array}
\right) = \hat{C}_6^{-1}
\left(
  \begin{array}{cccc}
      & e^{-{\rm{i}}\pi/6} &   &   \\
    e^{-{\rm{i}}\pi/6} &   &   &   \\
      &   &   &  e^{{\rm{i}}\pi/6} \\
      &   & e^{{\rm{i}}\pi/6} &   \\
  \end{array}
  \right)   \left(
  \begin{array}{c}
    | M_1, \hat{C}_6 \bm{q}, A, \uparrow \rangle \\
    | M_1, \hat{C}_6 \bm{q}, B, \uparrow \rangle \\
    | M_1, \hat{C}_6 \bm{q}, A, \downarrow \rangle \\
    | M_1, \hat{C}_6 \bm{q}, B, \downarrow \rangle \\
  \end{array}
\right),
\end{eqnarray*}
and
\begin{eqnarray*}
\left(
  \begin{array}{c}
    | M_3, \bm{q}, A, \uparrow \rangle \\
    | M_3, \bm{q}, B, \uparrow \rangle \\
    | M_3, \bm{q}, A, \downarrow \rangle \\
    | M_3, \bm{q}, B, \downarrow \rangle \\
  \end{array}
\right) = \hat{C}_6
\left(
  \begin{array}{cccc}
      & e^{{\rm{i}}\pi/6} &   &   \\
    e^{{\rm{i}}\pi/6} &   &   &   \\
      &   &   &  e^{-{\rm{i}}\pi/6} \\
      &   & e^{-{\rm{i}}\pi/6} &   \\
  \end{array}
  \right)   \left(
  \begin{array}{c}
    | M_1, \hat{C}_6^{-1} \bm{q}, A, \uparrow \rangle \\
    | M_1, \hat{C}_6^{-1} \bm{q}, B, \uparrow \rangle \\
    | M_1, \hat{C}_6^{-1} \bm{q}, A, \downarrow \rangle \\
    | M_1, \hat{C}_6^{-1} \bm{q}, B, \downarrow \rangle \\
  \end{array}
\right).
\end{eqnarray*}
where the phases $e^{\pm {\rm{i}}\pi/6}$ in the expression for $M_2$ ($M_3$) come from rotation about $\hat{z}$-axis by $\mp \pi/3$ on the spin states while the $\hat{C}_6$ is an operator of the rotation about $\hat{z}$-axis by $\pi/3$ on the vectors in real space or momentum space. For simplicity, we use $U_{21}$ and $U_{31}$ to represent the above two square matrices. As a result, the Hamiltonian around $M_2$ can be obtained as follows
\begin{eqnarray*}
\langle M_2, {\bm{q}}, i | H | M_2, {\bm{q}}, j \rangle = U_{21}(i,m)^{\prime} \langle M_1, \hat{C}_6 {\bm{q}}, m | \hat{C}_6 H
\hat{C}_6^{-1} | M_1, \hat{C}_6 {\bm{q}}, n \rangle U_{21}(j,n).
\end{eqnarray*}
It is noteworthy that the real-space Hamiltonian $H(t_{\rm{I}},t_{\rm{IR}},\lambda,\varphi)$ shown in Eq.~\eqref{MLTBH} is not an invariant under the $\hat{C}_6$ or $\hat{C}_6^{-1}$ operation, which not only interchanges the A and B sublattices making $t_{\rm{IR}}$ become $-t_{\rm{IR}}$ since $\mu_{ij}$ has opposite signs for A and B sublattices but also changes the orientation of magnetization $\varphi$ to $\varphi \mp \pi/3$. This indicates that
\begin{eqnarray*}
H_{{M_2}}(\bm{q},i,j) = U_{21}^{\rm{T}\dag}(i,m) H_{\rm{M1}}(-t_{\rm{IR}},\hat{C}_6\bm{q},m,n,\varphi+\pi/3) U_{21}^{\rm{T}}(n,j)
\end{eqnarray*}
\begin{eqnarray*}
H_{{M_2}}(\bm{q})=U_{21}^{\rm{T}\dag} H_{\rm{M1}}(-t_{\rm{IR}},\hat{C}_6\bm{q},\varphi+\pi/3)U_{21}^{\rm{T}}.
\end{eqnarray*}
The effective Hamiltonian around $M_3$ can be obtained in a similar manner, which can be expressed as follows
\begin{eqnarray*}
H_{{M_3}}(\bm{q})=U_{31}^{\rm{T}\dag} H_{\rm{M1}}(-t_{\rm{IR}},\hat{C}_6^{-1}\bm{q},\varphi-\pi/3)U_{31}^{\rm{T}}.
\end{eqnarray*}

With these $4\times 4$ Hamiltonians obtained by expanding the momentum around three $M$ points, we can further derive the effective Hamiltonian of the low-energy two bands by using the second-order perturbation theory.\cite{QAHE_G_Qiao_12} We still take $M_1$ as an example. For $\delta \lambda \ll t$ with $\delta \lambda = \lambda - t $, the Hamiltonian exhibited above can be divided into two parts:
\begin{eqnarray}
H_{{M_1}}(\bm{q})  &=& H_{{M_1}}^0 + H_{{M_1}}^{\prime} \nonumber \\
&=& \left(
  \begin{array}{cccc}
     0 & te^{\frac{{\rm{i}}4\pi}{3}} & \varphi & 0 \\
     te^{-\frac{{\rm{i}}4\pi}{3}} & 0 & 0 & \varphi  \\
   \varphi^*  & 0 & 0 & te^{\frac{{\rm{i}}4\pi}{3}} \\
    0 & \varphi^* & te^{-\frac{{\rm{i}}4\pi}{3}} & 0 \\
  \end{array}
\right) +
\left(
  \begin{array}{cccc}
    \lambda_{\rm{I}} & -\delta \eta & \delta \lambda_{\rm{RA}} & 0 \\
    -\delta\eta^* & -\lambda_{\rm{I}} & 0 & \delta \lambda_{\rm{RB}} \\
    \delta \lambda_{\rm{RA}}^* & 0 & -\lambda_{\rm{I}} & -\delta \eta \\
    0 & \delta \lambda_{\rm{RB}}^* & -\delta \eta^* & \lambda_{\rm{I}} \\
  \end{array}
\right)
\label{H1eff4}
\end{eqnarray}
where $H_{{M_1}}^0$ is the dominated part contributed from the major terms of both nearest-neighbor hopping and in-plane magnetization with $\varphi=te^{-{\rm{i}}\varphi}$. $H_{{M_1}}^{\prime}$ is the perturbation part originating from SOCs and minor terms of both hopping energy and magnetization with $\delta \lambda_{\rm{RA, RB}} = \delta\lambda e^{-{\rm{i}}\varphi} \mp t_{\rm{IR}} (3q_y-{\rm{i}}q_x)$ and $\delta \eta = - t(2{\rm{i}}q_y/\sqrt{3} - q_x^2/4 + q_y^2/12)e^{\frac{{\rm{i}}4\pi}{3}}$.
The main part $H_{{M_1}}$ can be easily diagonalized with eigenenergies being $\{ 0,0,-2t,2t\}$ and the corresponding eigenfunctions being $\{\xi_{+}\otimes \zeta_{-}, \xi_{-}\otimes \zeta_{+},\xi_{-}\otimes \zeta_{-},\xi_{+}\otimes \zeta_{+} \}$ where $\xi_{\pm}=1/\sqrt{2}(\pm e^{\frac{{\rm{i}}4\pi}{3}} |A\rangle + |B\rangle)$ and $ \zeta_{\pm}=1/\sqrt{2}(\pm e^{-{\rm{i}}\varphi} |\uparrow\rangle + |\downarrow\rangle)$. One can find that the first two basis functions correspond to the low-energy part while the other two basis functions are the high-energy part. After a unitary transformation, we can express the Hamiltonian of Eq.~\eqref{H1eff4} in the basis of these four functions with well separated low- and high-energy parts. Then we can obtain the low-energy two-band effective Hamiltonian by using the second-order perturbation theory introduced in Ref.~[\onlinecite{QAHE_G_Qiao_12}]. In the vicinity of  $M_1$, the effective Hamiltonian can be derived to be:
\begin{eqnarray}
h_{M_1}(\bm{q}) = m\sigma_z + a \sigma_x + b \sigma_y
\label{Heff2M1}
\end{eqnarray}
where $m=-\delta\lambda+(3q_y^2-q_x^2)/4$, $a=4q_xt_{\rm{I}}$, and $b=t_{\rm{IR}}(3q_y\sin \varphi-q_x\cos \varphi)$ with $\sigma_{x,y,z}$ being Pauli matrices. These two basis functions are mainly contributed from $\{\xi_{+}\otimes \zeta_{-}, \xi_{-}\otimes \zeta_{+} \}$.
By employing similar procedure, the effective Hamiltonian of $M_2$ and $M_3$ can be obtained that can be written as below
\begin{align}
\nonumber h_{M_2}(\bm{q})&= [\delta\lambda-t (q_x^2/2 + q_x q_y\sqrt{3}/2)] \sigma_z + 4(-q_x/2+\sqrt{3}q_y/2) t_{\rm{I}} \sigma_x \nonumber \\
& -[(2q_x+\sqrt{3}q_y)\cos\varphi+\sqrt{3}q_x\sin\varphi] t_{\rm{IR}} \sigma_y \\
h_{M_3}(\bm{q}) &=  [\delta\lambda-t (q_x^2/2 - q_x q_y\sqrt{3}/2)] \sigma_z + 4(-q_x/2-\sqrt{3}q_y/2) t_{\rm{I}} \sigma_x \nonumber \\
& + [(2q_x-\sqrt{3}q_y)\cos\varphi-\sqrt{3}q_x\sin\varphi] t_{\rm{IR}} \sigma_y.
\end{align}
We find that through a coordinate transformation that changes $\bm{q}$ to $\hat{C}_6^{-1}\bm{q}$ ($\hat{C}_6 \bm{q}$) in $h_{M_2}$ ($h_{M_3}$), these two low-energy two-band effective Hamiltonian can be expressed as follows by using similar parameters in Eq.~\eqref{Heff2M1}
\begin{eqnarray}
\nonumber h_{M_2}(\bm{q})&=& - [m\sigma_z + a \sigma_x + b(\varphi+\pi/3) \sigma_y] \\
h_{M_3}(\bm{q}) &=& - [m\sigma_z + a \sigma_x + b(\varphi-\pi/3) \sigma_y].
\end{eqnarray}
\end{widetext}

\end{appendix}

\end{document}